# Conductive Scaffolding for Neural Tissue Regeneration: 3D Bridging with Two-Photon Fabrication

Vladimir Osipov*, Aminah Jawed , Petro Lutsyk, David J. Webb and Antonio Fratini

Aston University, B4 7ET, Birmingham, UK

**Abstract.**
We report on a novel dual-structure scaffold fabricated using two-photon polymerisation (2PP), integrating electrically conductive and non-conductive regions within a single architecture, targeting neural tissue regeneration. Poly(ethylene glycol) diacrylate (PEGDA) was combined with 20 nm gold nanoparticles to create a conductive microstructure, encapsulated within a biocompatible PEGDA lattice designed to support neuronal growth possibility after neural injuries. Electrical resistance measurements confirmed the feasibility of integrating conductive pathways into precision-fabricated microarchitectures.

*Vladimir Osipov, v.osipov@aston.ac.uk

## 1 Introduction.

In the realm of neural tissue engineering, conductive scaffolds represent a critical innovation for brain tissue regeneration. These specialised structures go beyond traditional tissue support, incorporating electrical conductivity to mimic the complex signalling environment of neural networks. The primary challenge lies in creating a scaffold that not only provides structural support for neural cells but also facilitates electrical signal transmission, a key requirement for functional neural tissue reconstruction [1].

Conductive scaffolds typically integrate materials with specific electrical properties, such as carbon nanotubes, graphene, or doped polymers, that can effectively conduct electrical signals while maintaining biocompatibility [2,3]. The goal is to create a three-dimensional environment that closely replicates the natural neural microarchitecture, supporting cell adhesion, migration, and functional reconnection. The design of these scaffolds must address multiple critical parameters: electrical conductivity, mechanical strength, surface topography, and biochemical compatibility [4]. Each of these factors plays a crucial role in guiding neural cell behaviour, promoting axonal growth, and potentially facilitating neural circuit reconstruction in cases of traumatic brain injury or neurodegenerative conditions.

For more than half a century, various fields including tissue engineering and regenerative medicine have demanded high-resolution, well-defined 3D microstructures to create effective cellular scaffolds [5]. While photolithography has long dominated microfabrication, it and its derivatives are fundamentally two-dimensional, severely limiting their ability to create the sophisticated 3D environments necessary for proper tissue regeneration [6].

Traditional 3D scaffold fabrication approaches each have significant limitations for tissue engineering applications. The LIGA process can create simple 3D structures but struggles with the complex architectures needed to mimic natural tissue environments [7]. Self-assembly faces difficulties in producing the specific, controlled structures necessary for guiding cellular growth. Common 3D printing techniques rely on layer-by-layer approaches, which imposes geometric constraints and typically cannot achieve the micron-level resolutions necessary to control individual cellular behaviour [8]. Traditional techniques for producing 3D scaffolds, including phase separation, particulate leaching, solvent casting, and freeze-drying, typically achieve resolutions of only tens of microns, which is insufficient for controlling individual cellular interactions. While electrospinning can produce nano-scale fibres for tissue scaffolds, the dense accumulation often hinders cell infiltration and proper tissue formation. These conventional methods struggle to accurately reproduce the complex architectures found in natural tissues.

Two-photon polymerization (2PP), developed in 1997, overcomes these limitations [9,10]. TPP works through optical nonlinear absorption typically using a near-infrared femtosecond laser to induce polymerization in photosensitive materials [11]. The photo initiator molecule simultaneously absorbs two photons to trigger local free-radical polymerization within a precisely defined focal volume, allowing for true 3D fabrication of tissue scaffolds without the geometric constraints of layer-by-layer methods [12]. TPP offers exceptional resolution capabilities particularly suited for tissue engineering, achieving sub-diffraction limited resolution with structures featuring details of several hundred nanometres [13]. This precision is critical for creating the fine features necessary to guide cell adhesion, migration, and differentiation within regenerative scaffolds. Over the past two decades, TPP has evolved into a practical method widely employed for creating sophisticated biomedical scaffolds that can effectively support tissue growth and regeneration [14]. In tissue engineering applications, TPP has become particularly valuable for creating scaffolds that accurately mimic the microenvironments of natural extracellular matrices [15].

The 2PP fabrication laser technique [16] allows for direct fabrication of designed micro-scaffolds and porous-like micro-structures on glass substrates and around the optical fibre facet with precision down to 100 nm: this is a radical and flexible concept that is not achievable by established fabrication techniques, which leads to a range of novel device possibilities. From another side, 2PP is suitable to fabricate a set of compact optical components (lenses, waveguides, gratings etc) and can be used to create the micro-fluidic structure of a device incorporating precision mechanical components such as locks and shutters. Earlier it was used for fabrication of biomedical micro-optics like three-focus intraocular lens [17], and intra-vascular pressure sensing [18].

## 2 Methods

*2.1 Design & Experimental*

**Design:** Dual-structure concept modelled in SolidWorks. Because of time limitations for this study only the most challenging conductive core was fabricated and tested.

To preserve resolution and maintain compatibility with the system's working distance limitations final scaffold dimensions were chosen approximately 110 µm in width, 62 µm in depth, and 36 µm in height. The beams thickness was 4 x 4 µm, smaller square holes 8 x 8 µm, central hole 15 x 15 µm. With a scaffold height of 36 µm, basement height of 10 µm and individual layer thickness of 0.5 µm, the 2PP printing process required 72 layers for full fabrication of the structure.

**Resins:** early trials used PEGDA + 20 nm AuNPs + Irgacure 819; final prints used PEGDA + 20 nm AuNPs with LAP (lithium phenyl-2,4,6-trimethylbenzoylphosphinate). Example mixes reported: 4 mL PEGDA + 2 mL AuNPs + 0.2 g Irgacure 819; then 2 mL PEGDA + 1 mL AuNPs +~0.1gLAP.

**Substrates:** 18 mm gold-coated glass slides with a scalpel scratch (width ~10 µm) to create an open circuit; prints were written across the gap and conductivity assessed with a multimeter.

*2.2. 2PP setup*

Fabrication was realized using LaserNanoFab 2PP fabricating setup, based on a femtosecond laser source Chromacity Spark 1040HP (pulse duration 130 fs, repetition rate 100 MHz) system at 520 nm with the Olympus 40X objective. Inverted droplet mounting was used to minimise meniscus aberrations.

Laser parameters: Initial PEGDA–AuNP tests polymerised at 4–5 mW on glass substrate; printing onto gold required a higher dose with stable polymerisation at about 22 mW at 520 nm. A representative build used 0.5 µm layer thickness for a 36 µm structure (72 layers) with a total fabrication time of about 22 min 48 s.

Fabrication parameters: Conductive core geometry employed cylindrical struts. Zero-offset was set a few microns into the substrate to secure surface adhesion. Development was in deionised water for about 5 minutes, then air or nitrogen dry, followed by resistance measurement across the bridged scratch.

## 3 Results

Conductive PEGDA–AuNP microstructures successfully bridged gold slide gaps, demonstrating structural integrity and measurable dimensions. After scaffold printing, resistance was restored in two samples, with values ranging from 1.3 Ω to 100 Ω, confirming that scaffold enabled electrical conduction across the gap (Fig.1).

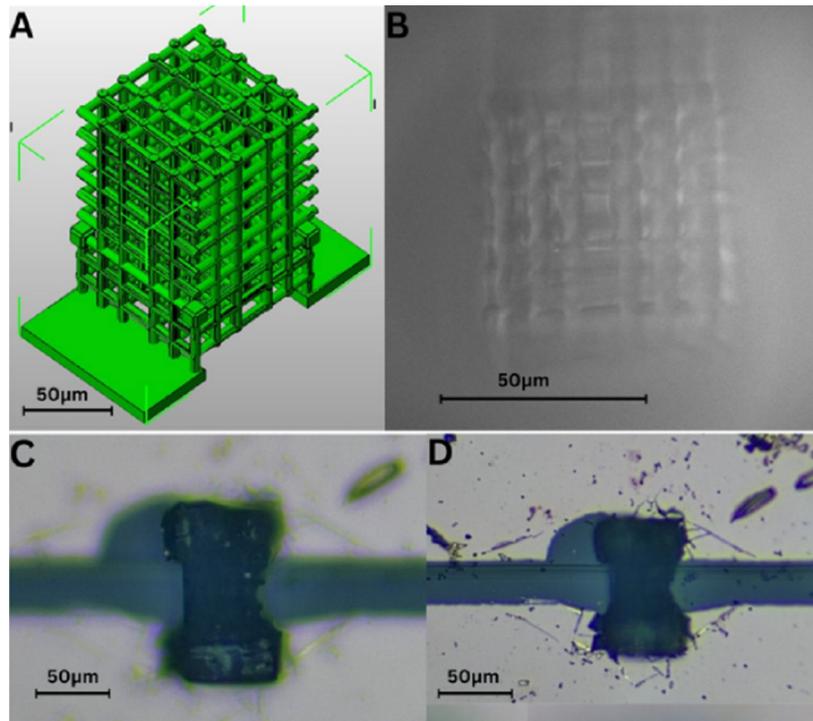

Fig.1. A: Computer design of the scaffold. B: Optical microscopy images of PEGDA + 20 nm AuNP conductive microstructures in polymer just after 2PP-fabrication. C, D: Optical microscopy images of the conductive microstructures in air after development at high/low focus.

**We carried out AFM study**, exploring 2PP fabricated conductive microstructures in air after development.

**Equipment Used:** Bruker Dimension Icon® Atomic Force Microscope with inspectable area 180x150x40 mm3, noise < 30 pm RMS, 90x90x10 µm3 scan size. Probes used: RTEPSA-300-125, RFESPA-75, SCANASYST-AIR-HPI. The view of the cross-sectional height analysis along the gap (between metal stripes) is shown in Fig.2.

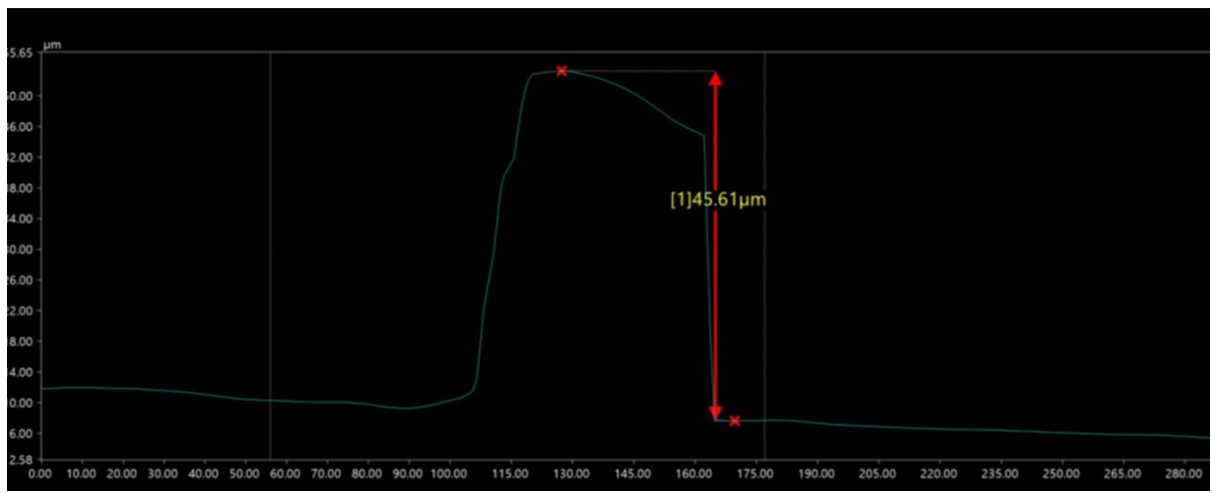

Fig.2. 3D optical profilometry and cross-sectional height analysis (~45.6 µm) showing the printed structure's geometry and alignment across the gap.

## Conclusions

Conductive PEGDA–AuNP microstructures were fabricated via two-photon polymerisation, successfully bridging microscale gaps on gold-coated slides with defined geometry and stable adhesion. The results demonstrate the feasibility of producing precise conductive pathways in microscale architectures, forming a strong basis for future integration into dual-structure scaffolds for neural tissue regeneration, where both structural support and electrical functionality are required. This work provides a proof-of-concept that can be further developed towards in vitro testing and eventual clinical translation.

## Disclosures

The authors declare that there are no financial interests, commercial affiliations, or other potential conflicts of interest that could have influenced the objectivity of this research or the writing of this paper.

## Code and Data Availability

All datasets used in this work are publicly available through the corresponding references listed. In the meantime, specific inquiries can be directed to the corresponding author.

## Acknowledgements.

We would like to gratefully acknowledge the LEVERHULME TRUST for support of this work, Research Project Grant RPG-2022-334.

## References.

**Vladimir Osipov** has been a researcher with Stepanov Institute of Physics (SIP), Belarus since 1976, Marie-Curie Fellow with the Laser Zentrum Hannover (LZH), Germany and Research Associate with Aston University, UK since 2017. He received his PhD in quantum electronics from SIP (1997). He has been awarded nine patents and has more than 75 technical publications. His interests include two-photon polymerization, micro-optics/micro-devices design and laser fabrication.

**David J. Webb** joined Aston University as Reader in Photonics in 2001, was promoted to Professor in 2012 and awarded a 50th Anniversary Chair in 2016. Prior to joining Aston he had spent 10 years as a Lecturer, then Senior Lecturer in the Physics Department at the University of Kent at Canterbury. He received BA in Physics (1982) at the University of Oxford and the PhD in Physics (1988) at the University of Kent. His research interests primarily encompass optical fibre sensing and devices, on which topics he has over 400 publications.

**Petro Lutsyk** received his PhD in physics and math at the Institute of Physics, National Academy of Sciences of Ukraine in 2008. After a decade of post-doctoral research projects in Poland, Ukraine, Italy, & UK, he joined Aston University as Marie-Curie Fellow in 2015, then becoming Lecturer in Electronic and Systems Engineering in 2018; promoted to Senior Lecturer in 2024. He is also associate member of the AIPT having multidisciplinary research experience with a proven track record (co-authored 48 peer-review scientific articles) in thin film device engineering for electronics & photonics, nanotechnology & materials science.

**Antonio Fratini** joined Aston University in 2022, as a Reader. He is a bioengineer with a vocation of improving clinical outcomes and patient participation to a more efficient, informed, and inclusive healthcare. He graduated with an MSc in Electronics Engineering from University of Naples Federico II in 2005 (with a specialisation in Medical Instrumentation), followed by a PhD at the University of Bologna field of Bioengineering in 2009 and a further PhD in Healthcare systems management in 2016. His research interests lie in the areas of biomedical data processing (imaging, segmentation, modelling, AI), medical devices (biosensing and diagnostics) and Healthcare processes (technology assessment and management).